# ON THE STATISTICAL LAW OF LIFE

N. M. Pugno
Department of Structural Engineering, Politecnico di Torino,
Corso Duca degli Abruzzi 24, 10129, Italy

**Abstract**
In this paper we derive a statistical law of Life. It governs the probability of death (or complementary of survival) of the living organisms. We have deduced such a law coupling the widely used Weibull (1951) statistics, developed for describing the distribution of the strength of solids, with the "universal" model for ontogenetic growth only recently proposed by West and co-authors (2001). The main idea presented in this paper is that cracks can propagate in solids and cause their failure as sick cells in living organisms can cause their death. Making a rough analogy, living organisms are found to behave as "growing" mechanical components under cyclic, i.e., fatigue, loadings and composed by a dynamic evolutionary material that, as an ineluctable fate, deteriorates. The implications on biological scaling laws are discussed. As an example of application, we apply such a statistical law to large data collections on human deaths due to cancer of various types recorded in Italy: a relevant agreement is observed.

**Introduction: Weibull's and West et al.'s approaches**
Weibull Statistics (Weibull, 1951) describes the statistical distribution for the strength of solids. Weibull derives the cumulative probability of failure $P_f$ for a structure of volume $V$ and subjected to a uniaxial local stress $\sigma(\bar{x})$, a function of the position vector $\bar{x}$, as $P_f = 1 - \exp\left[-\int_V \left(\frac{\sigma(\bar{x})}{\sigma_{0V}}\right)^m dx^3\right]$, where $\sigma_{0V}$ and $m$ are respectively Weibull's scale and shape parameters. The former presents anomalous physical dimension and is related to the mean value of the distribution, whereas the latter is dimensionless and related to its standard deviation. If a threshold value $\sigma_{th}$ for the failure stress is observed, the substitution $\sigma(\bar{x}) \rightarrow \sigma(\bar{x}) - \sigma_{th}$ into the expression for $P_f$ has to be considered. For structures with surface-flaws that are dominating with respect to volume-flaws, e.g., at small size-scales where the ratio surface over volume tends to diverge, the surface $S$ of the structure must replace its volume $V$.

For uniform uniaxial stress $\sigma$, thus coincident with the applied load, e.g., the case of a fiber in tension, the probability of failure $P_f$ becomes:

$$P_f = 1 - \exp\left[-V\left(\frac{\sigma}{\sigma_{0V}}\right)^m\right] \qquad (1)$$

The presence of the volume (or of the surface) in eq. (1) is crucial: Weibull assumed a number of potential "critical" defects in the structure, as statistically proportional to its volume $V$ (or to its surface $S$). The impact on Materials Science of this pioneer Weakest

Link Theory, a by-product of the general Statistics of the Extremes (Gumbel, 1958), has been tremendous.

This suggests to make an analogy between Mechanics and Biology: we can treat a living organism as a mechanical component, the pre-existing defects in the structure corresponding to biological defects in the organism, e.g., potential sick cells. Correspondingly, a fully crack propagation will cause the failure of the specimen as well as a fully propagation of the biological defect will cause the death of the organism. Even if Weibull Statistics is currently applied in Biology (Cherkasov et al., 2004; Krishnan et al., 2004; Wu et al., 2004; Ng et al., 2004; Miceli et al., 2004; May et al., 2004; Weston et al., 2004; Molina et al., 2004) we demonstrate in this paper the importance of considering dynamic Weibull parameters, thus a new statistics.

Starting from this analogy, there are three questions that we have to answer before trying to develop a general statistical law of Life: (1) which parameter plays the role of the stress in Life? (2) is the Weibull's modulus $m$ a constant for a living organism or it presents a dynamic evolution as the organism itself? Finally, (3) is the probability of death of a living organism related to its "growing" volume (or mass $M$), or to its surface or to what else?, and if the growing process plays a role, how we can take into account a "universal" growing of the living organisms?

The answer to question (1) seems to be unquestionable: it is the time $t$. Regarding this point we note that the Weibull Statistics, in which the stress is replaced by the time, is widely used also for describing the statistics of the times to failure for mechanical components under cycling loads. Thus, living organisms would behave as mechanical components under fatigue. A cycle can be considered as a biological characteristic period, as for example a day. We will demonstrate that the answer to question (2) is that the Weibull's modulus is in general evolutionary as the living organism itself. To give an answer to question (3) would require to describe in a "universal" manner the growing process of the living organisms, i.e., the mass-time dependence $M(t)/M_\infty$, regardless to their asymptotic mass $M_\infty$, that spans 21 orders of magnitude, from microbes to whales.

Regarding this point, we consider the general model for ontogenetic growth recently proposed by West and co-authors (2001). They demonstrated that the growth of the living organisms can be universally described by the following mass/energy balance:

$$\frac{dM}{dt} = a\, M^p \left[1 - \left(\frac{M}{M_\infty}\right)^{1-p}\right] \tag{2a}$$

where $a$ is a constant related to the metabolic rate of the living organism. West et al. (2001) argue that a value of $p=3/4$ has to be considered to describe natural fractal-like energy distribution networks; however this hypothesis can be also relaxed (Guiot et al., 2003): it would correspond to a different dimension of the fractal set, as demonstrated in a different context by Carpinteri and Pugno (2002). According to these authors, a value of $p$ comprised between $2/3$ and $1$ is expected; $p=2/3$ would describe surface dominated energy supply mechanisms, whereas for $p=1$ volume effects prevail. The universal growth predicted by the previous equation is:

$$r = 1 - e^{-\tau}, \text{ with } r = (M/M_\infty)^{1-p} \text{ and } \tau = a(1-p)M_\infty^{p-1}t - \ln(1-r_0) \qquad (2b)$$

where $r_0 = (M_0/M_\infty)^{1-p}$ with $M_0 = M(t=0)$. Thus, the function $M(t)$ is now "universally" quantified. Usually, only the parameter $a$ is unknown. However, we could estimate it by the time at which the organism has reached a conventional ratio $M/M_\infty$. For example if $t_{97\%}$ is defined by $M(t_{97\%})/M_\infty = 0.97$, considering for human individuals $t_{97\%} \approx 25$ years, $M_0 \approx 3$ Kg, $M_\infty \approx 80$ Kg, and $p=3/4$ would correspond to $a \approx 2\,\text{Kg}^{1/4}/\text{year}$. Coupling the Weibull's and West at al.'s approaches, on the basis of the analogy previously presented, a statistical law of the Life can be formulated, as described in the next Section.

**Towards the statistical law of Life**
Coupling the Weibull's and West at al.'s approaches, we formulate a statistical law of the Life giving the probability of death $P_D$, or of survival $P_S$, as suggested by eqs. (1) and (2):

$$P_D = 1 - P_S = 1 - \exp\left[-kM^\gamma(t)\left(\frac{t}{t_0}\right)^{m(t)}\right] \qquad (3a)$$

where $M(t)$ is given by eq. (2b), $m(t)$ is an evolutionary modulus, $k$ is a constant with anomalous physical dimension and $t/t_0$ is a dimensionless time, defined by arbitrarily fixing $t_0$, e.g., $t_0 = 1$ year. We have considered $M^\gamma$ instead of $M$ since considering the surface instead of the volume (or mass) in eq. (1) would correspond to replace $M$ with $M^{2/3}$, if individuals belonging to a specified organism family are assumed to be geometrically self-similar. Thus, for generality, a constant $\gamma$ has to be introduced. To take into account a threshold time $t_{th}$, we could replace $t \rightarrow t - t_{th}$ in the explicit time dependence of eq. (3a), as suggested by the Weibull's treatment. Note that setting $\gamma = p$ would correspond to assume a direct dependence from the average resting metabolic rate of the organism, derived as proportional to $M^p$ by West et al. (2001), that assumed $p = 3/4$. By setting $\gamma = 0$ we could model an explicit independence from the mass, so that the growing process would be only implicitly modeled by the dynamic evolution of $m(t)$. Negative value of $\gamma$ would reveal an inverse mass dependence, unrealistic in Mechanics but perhaps plausible in Biology. Formally, the parameter $\gamma$ allows one to consider any interesting function into eq. (3a): if for example we want to consider the growing rate $dM/dt$ instead of $M^\gamma$, this would correspond to $\gamma = \ln(dM/dt)/\ln M$. This would imply a probability of finding potential critical cells as proportional to the growing rate that, according to eq. (2b), presents a maximum for $M/M_\infty = p^{1/(1-p)}$. The proposed mechanical/biological analogy is summarized in Table 1.

## Scaling laws in Mechanics and Biology

Let us consider structures composed by a given material but having different size scales. It is well-known that eq. (1) implies a scaling for their nominal failure stresses according to $\sigma_f^m V \approx \text{const}$ (Weibull, 1951); this can be deduced by setting $P_f = 0.63 = \text{const}$, 0.63 being the value of the probability defining the nominal stress. The size-scale effect $\sigma_f \propto V^{-1/m}$ suggests that *smaller is stronger* as nowadays well-known in Mechanics (e.g., see Carpinteri and Pugno, 2004). This represents the reason of why nanostructures, such as nanotubes or nanowires, show tremendous high strength: it is a consequence of the reduced probability of finding critical defects in their minimal volume (see Pugno and Ruoff, 2004).

Similarly, a scale-time effect is deduced according to eq. (3a) as $kM^\gamma(t_D)(t_D/t_0)^{m(t_D)} \approx \text{const}$. Since statistically $M(t_D) \propto M_\infty$, a biological time scaling law $t_D \propto k^{-1} t_0^{m(t_D)} M_\infty^{-\gamma/m(t_D)}$ between $t_D$ and $M_\infty$ is finally derived. Consequently, some information on the parameter $\gamma$ can in principle be obtained by matching well-known biological scaling laws with our prediction. For example, generalizing for a generic value of *p* the biological scaling laws reported by Brown and West (2000; that assume *p*=3/4), metabolic rate scales as $M_\infty^p$, radii of mammalian aortas and tree trunks as $M_\infty^{p/2}$, mammalian heart and respiratory rates as $M_\infty^{p-1}$, as well as circulation times for blood of mammals and sap of trees scale as $M_\infty^{1-p}$: such a scaling is expected for most biological times, including those of respiratory and cardiac cycles and gestation, postembryonic development and life span. Accordingly, $t_D \propto k^{-1} t_0^{m(t_D)} M_\infty^{-\gamma/m(t_D)} \propto M_\infty^{1-p}$ gives an additional information on the constants involved in eq. (3a), and in particular seems to suggest slightly negative values for $\gamma$. However, as a consequence of the small value of the exponent $1-p$, the time-scale effect seems to be negligible in this context, i.e., $\gamma \approx 0$, as derived in the next Section by the data treatment.

Thus, as $\sigma_f \propto V^{-1/m}$ in Mechanics, $t_D \propto M_\infty^{1-p}$ in Biology. The parameter *p*, as previously discussed, is in general expected to be comprised between 2/3 and 1 (Carpinteri and Pugno, 2002): this is confirmed by the experimental observations that reveal *p* not identical but close to 3/4 for all the analyzed living organism families (Brown and West, 2000). For details on this point see Guiot et al. (2005).

## Deterioration as ineluctable fate

Let us consider the public domain data bank of the Italian National Institute of Statistics (ISTAT, see the web page reported in the references). In particular we refer to the Tables of mortality due to cancers of various types recorded in Italy. The age of the individuals deaths are divided into time-intervals, 1-4, 5-9, ..., *i*-(*i*+4), ..., 75-79 years. For each time-interval *i* the number $N_i$ of the observed deceases, for a specified year and in Italy, is reported. We consider the deaths related to the time-interval *i* as arising at its mean value $t_i$ and we calculate the cumulative probability of death as $P_D(t_i) = \dfrac{\sum_{j=1}^{i} N_j - 1/2}{N}$

(Johnson, 1983), where $N$ is the total number of deceases (i.e., $N = \sum_j N_j$): thus, we are here considering relative probabilities. Alternatively, we could also refer to the total number of individuals $N = N_{TOT}$, that is the population in Italy in the investigated year: in this case the probabilities would be absolute. In Figure 1 the data related to the year 1990 for males and females, deceased as a consequence of cancer of various types, are treated by assuming $m$=const in eq. (3a), and alternatively $\gamma = 1$ (explicit mass dependence), $\gamma = \ln(\mathrm{d}M/\mathrm{d}t)/\ln M$ (explicit growing rate dependence) or $\gamma = 0$ (explicit independence from the growing process). We have fixed $t_0 = 1 \text{year}$.

Note that eq. (3a) could be put in the following form:

$$\ln(\ln(1/(1-P_D))) - \gamma \ln(M) = \ln(k) + m \ln(t/t_0) \tag{3b}$$

Thus, if $m$=const and an explicit linear dependence ($\gamma = 1$) from the mass are reasonable hypotheses, in the diagram of Figure 1, the squares should fall on a straight line, as predicted by eq. (3b). This clearly does not happen. At the same manner the triangles in Figure 1 should fall on a straight line if $m$=const and independence from the growing process ($\gamma = 0$) are reasonable assumptions. Again, this is not the case. The same curved trend is observed for an explicit dependence from the growing rate ($\gamma = \ln(\mathrm{d}M/\mathrm{d}t)/\ln M$, $m$=const, rhombs). A similar trend is observed also considering $\gamma = -1$. Such curved trends suggest that for human cancer the mass $M$ or its growing rate $\mathrm{d}M/\mathrm{d}t$ has not an explicit crucial role; thus, we can treat these data by setting $\gamma = 0$. On the other hand, the discrepancy from straight lines observed in Figure 1, emphasizes the importance of considering an evolutionary behaviour for $m$, related to the slopes of the curves in Fig. 1. Such evolution can be physically clarified noting that, according to reliability theory, a Weibull distribution with $m>1$ characterizes a life system that increasingly deteriorates. On the other hand, if the shape parameter is smaller then unity ($m<1$), there is a reliability growth as the failure rate of the system decreases with time (Cherkasov et al., 2004). Thus, since the slopes of the curves in Fig. 1 increase, the corresponding life systems tend, as an ineluctably fate, to deteriorate. The simplest assumption is to consider the linear dependence $m = t/t_m$, where $t_m$ is a biological characteristic time. Considering other hypotheses, such as $m \propto \sqrt{t}$ or $m \propto t^2$ would make the situation worse. The physical meaning of the parameter $t_m$ is clear: for $t < t_m$ no deterioration occurs, whereas for $t > t_m$ a catastrophic deterioration is expected. In these hypotheses, the more general eqs. (3a) and (3b) become:

$$P_D = 1 - P_S = 1 - \exp\left[-k\left(\frac{t}{t_0}\right)^{t/t_m}\right] \tag{4a}$$

$$\ln(\ln(1/(1-P_D))) = \ln(k) + (t/t_m)\ln(t/t_0) \tag{4b}$$

If Nature follows, for human cancer diseases, the statistical law of eq. (4a), the same data reported in Figure 1 should fall in the diagram of Figure 2 on a straight line, as suggested by eq. (4b). A noticeable agreement is observed (see the coefficient of correlation $R^2$), suggesting that such a statistic could represent a kind of general biological law. In addition, the best fit gives in a simple manner the two characteristic constants $k, t_m$ for the analysed case. We find $k^{-1} \approx 769.7$ and $t_m \approx 44.6 \text{years}$. About the physical interpretation of the former parameter, we note that the larger the value of $k$ the lower the toughness against the considered decease, as suggested by eqs. (4) considering $t$=const. The latter represents the time corresponding to the beginning of a catastrophic deterioration. This allows one to classify the considered living organism family against the considered kind of death in a simple and powerful manner: from such parameters, statistical biological predictions could be easily obtained by applying eq. (4a).

**Cancer data analysis**
We choose to treat other data on cancer deaths to further test the statistics of eq. (4a). In Figure 3 we still refer to the year 1990 but considering separately males and females. Females ($k^{-1} \approx 643.7$ and $t_m \approx 46.1 \text{years}$) are found to be slightly stronger (larger $k^{-1}, t_m$) than males ($k^{-1} \approx 513.8$ and $t_m \approx 43.5 \text{years}$) against cancer. In Figure 4 a comparison between the years 1974 ($k^{-1} \approx 246.3$ and $t_m \approx 48.8 \text{years}$) and 1984 ($k^{-1} \approx 192.9$ and $t_m \approx 54.9 \text{years}$) is reported. The influence of the time, related to the different cancer aggressiveness and available therapies, is clearly observed: catastrophic behavior has been successfully retarded ($t_m$ increases with time), even if cancer seems to become more aggressive ($k^{-1}$ decreases). In Figure 5 different cancers (colon, larynx, mamma, melanoma, lung, prostate, stomach) are treated separately for the year 1990, as observed in Piemonte, a region of Italy. In all these cases the statistical law of Life of eq. (3a) in its particular version of eq. (4a) shows an impressive agreement. In Table 2 the characteristic constants $k^{-1}, t_m$ are reported as statistically deduced from the best fits of the data in Figure 5, as a rigorous purpose of classification.

**Conclusions**
Summarizing, it is clear that the proposed statistical law of Life could represent an interesting tool for classifying and deducing statistical predictions on the natural deaths of living organisms, as here demonstrated for cancer in human individuals. Further investigations may reveal the necessity of considering the more general eq. (3a) rather then its simplified version of eq. (4a). As the Weibull Statistics can be applied for predicting the probability of failure of a given family of structures, the statistical law of Life can be applied for predicting the probability of death for a given family of living organisms. For example for human individuals, all the most important causes of deaths (e.g., car accidents, HIV virus, heart attack, etc.) could be investigated with the proposed method. The universally accepted importance of the Weibull Statistics for describing the strength of solids would suggest that the derived statistical law of Life may have an interesting role in Biology and Medicine. Clearly, Weibull Statistics alone ($m$=const) has been demonstrated to be in general unable to catch the reality. An evolutionary parameter

*m*, intrinsically related to the deterioration process of the living organisms, seems to be crucial for correctly describing the statistics of Life.

**Acknowledgement**
The author would like to thank A. Carpinteri, P. P. Delsanto and C. Guiot, as well as Diane Dijak for the english grammar supervision.

TABLE CAPTIONS

Table 1: Statistical analogy between Weibull approach applied to solids and the statistical law of the Life applied to living organisms.

Table 2: Characteristic constants $k, t_m/t_0$ ($t_0 = 1\,\text{year}$) from the best fits of the different cancer typologies analyzed in Figure 5.

FIGURE CAPTIONS

Figure 1: Eqs. (3) with *m*=const, applied to human individuals deceased as a consequence of cancer (of various types) in the year 1990 in Italy: triangles ($\gamma = 0$), squares ($\gamma = 1$) and rhombs ($\gamma = \ln(dM/dt)/\ln M$); $M_0 \approx 3\,\text{Kg}$, $M_\infty \approx 80\,\text{Kg}$, $a \approx 2\,\text{Kg}^{1/4}/\text{year}$ (*p*=3/4).

Figure 2: Eqs. (4) applied to human individuals deceased as a consequence of cancer (of various types) in the year 1990 in Italy. The observation of a straight line seems to confirm the statistical law of Life here derived.

Figure 3: Eqs. (4) applied to human individuals deceased as a consequence of cancer (of various types) in the year 1990 in Italy. Males (squares) and females (triangles) are treated separately.

Figure 4: Eqs. (4) applied to human individuals deceased as a consequence of cancer (of various types) in the years 1974 (squares) and 1984 (triangles).

Figure 5: Eqs. (4) applied to human individuals deceased as a consequence of cancer of different types (colon, larynx, mamma, melanoma, lung, prostate, stomach) in the year 1990 in Piemonte, a region of Italy.

TABLES

| Probability of failure of a mechanical component | Probability of death of a living organism |
|---|---|
| 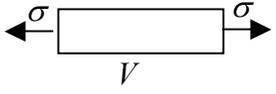 | 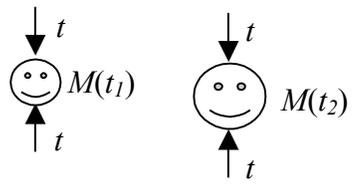 |
| Stress $\sigma$ | Time $t$ |
| $V$ or $S$ = time-independent | $M^\gamma(t)$ = time-dependent |
| $m$ = time-independent | $m(t)$ = time-dependent |

Table 1

|  | Colon | Larynx | Mamma | Melanoma | Lung | Prostate | Stomach |
|---|---|---|---|---|---|---|---|
| $t_m/t_0$ | 36.2 | 34.3 | 44.6 | 69.9 | 31.1 | 23.7 | 42.6 |
| $k^{-1}$ | 4665.2 | 2694.6 | 658.7 | 56.2 | 11185.8 | 11599.3 | 1627.0 |

Table 2

FIGURES

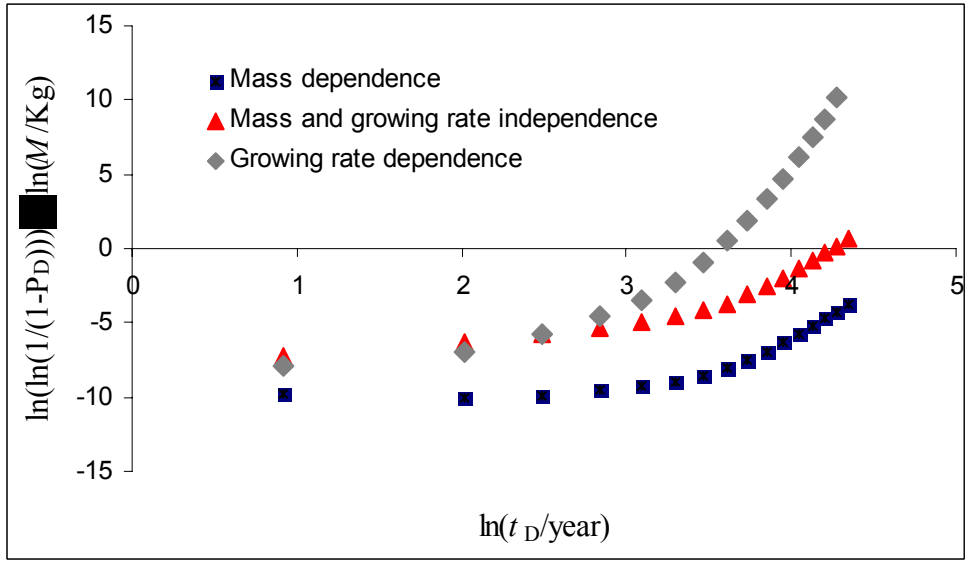

Figure 1

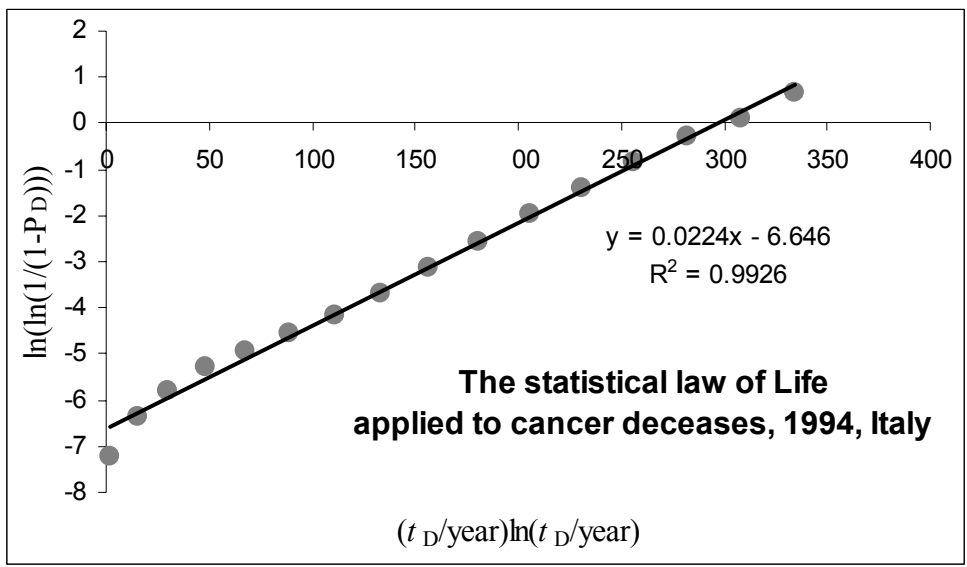

Figure 2

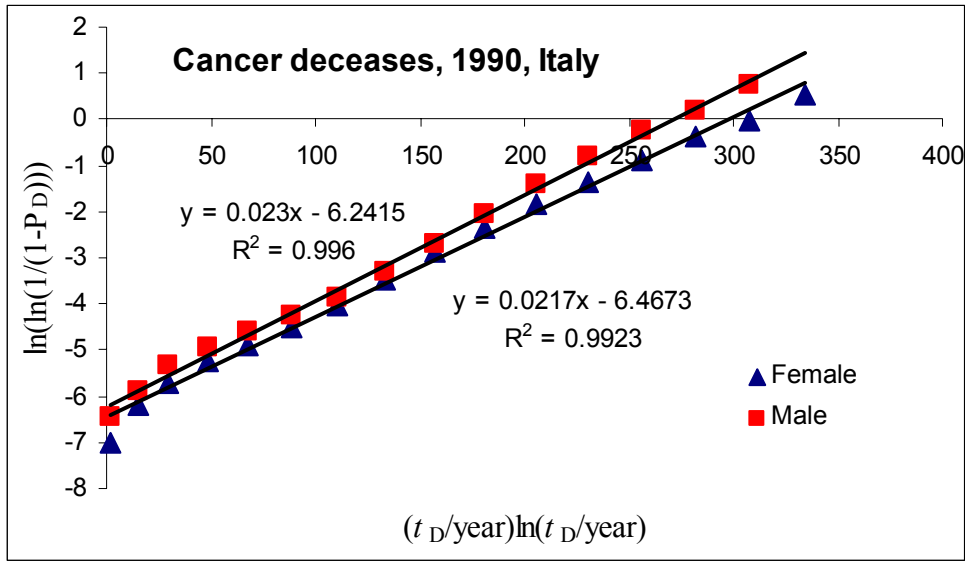

Figure 3

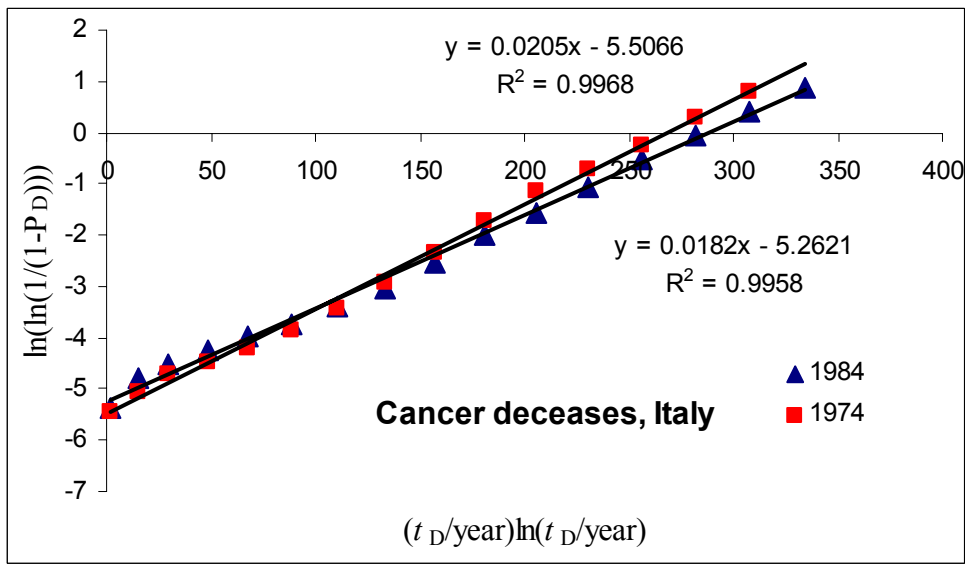

Figure 4

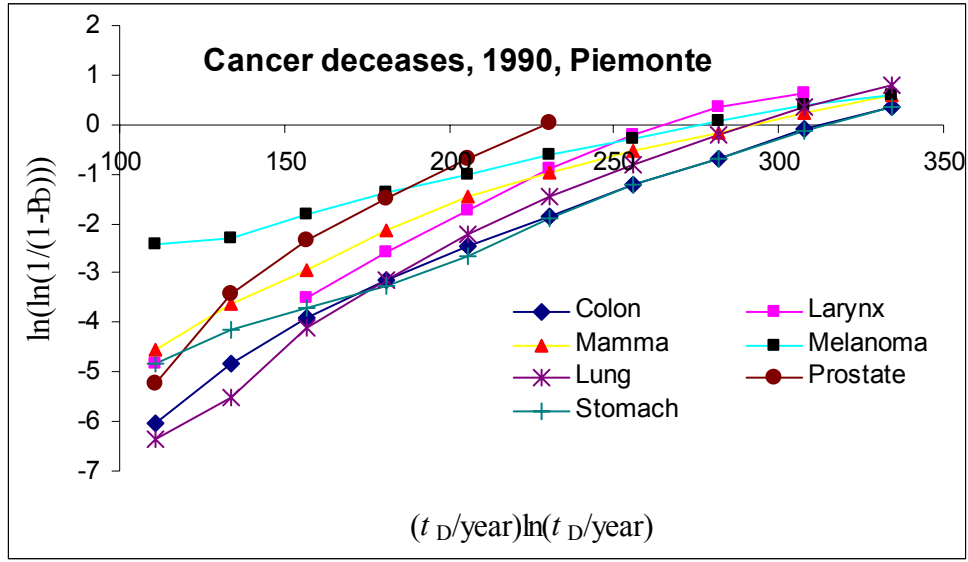

Figure 5